\documentclass[%
 reprint,
 superscriptaddress,
 amsmath,amssymb,
 aps,
]{revtex4-2}

\usepackage{graphicx}
\usepackage{dcolumn}
\usepackage{bm}
\newcommand{\RN}[1]
    {\MakeUppercase{\romannumeral #1}}

\usepackage{color}
\usepackage[normalem]{ulem}
\begin{document}

\preprint{APS/123-QED}

\title{A Theoretical Study of Cavity-modulated Topological Anderson Insulators}
\author{Yu-Cheng Shaw}
\thanks{pbture102@gmail.com}
\affiliation{Graduate Institute of Applied Physics, National Chengchi University, Taipei 11605, Taiwan\looseness=-1}

\author{Hsiu-Chuan Hsu}
\affiliation{Graduate Institute of Applied Physics, National Chengchi University, Taipei 11605, Taiwan\looseness=-1}

\author{J. S. You}
\affiliation{Department of Physics, National Taiwan Normal University, Taipei 11677, Taiwan\looseness=-1}




\date{\today}

\begin{abstract}
{Strong light-matter interaction has been demonstrated feasible for controlling phases of matter. 
In this work, the interplay with disorder is studied and rich phenomena are demonstrated.  
Specifically, the topological phases of the disordered longer-range Su–Schrieffer–Heeger (SSH) model coupled with cavity photons are studied numerically. 
It is found that cavity photons modify the hopping amplitudes, resulting in the change of phase transition boundaries, and disorder induced topological Anderson insulating (TAI) phases even in the presence of cavity photons. The critical disorder strength at the phase transitions, determined by localization lengths, can be modulated by cavity photons through the modified hopping amplitudes. Our work extends the study of cavity-coupled solid state systems to disordered lattices.  
}
\end{abstract}

\maketitle

\section{\label{sec:level1}Introduction}

The interest in controlling phases of matter with quantum light has grown recently, with cavity photons being one of the most popular light sources.
This allowed one to investigate a wide range of many-body and topological phenomena in cavity quantum electrodynamics~\cite{Dmytruk2022,Allard2023,Shaffer2024,Dmytruk2021,Feist2015,Hagenmuller2017,Masuki2024,Mochida2024,Dmytruk2024,Curtis2019,Leon2024,Schlawin2019,Sentef2018,Li2020,Bartolo2018,Masuki2023,Klinder2015,Landig2016,Kuroyama2024,Thomas2021,Orgiu2015,Krainova2020,Nagarajan2020,Bagliani2019,Appugliese2022,Knuppel2019,Bloch2022,Schlawin2022,Kockum2019,Hubener2021,Vidal2021}.
The theoretical methods for studying cavity-embedded materials have been developed.
To introduce coupling between bosonic and fermionic operators, attempts such as Peierls substitution~\cite{Dmytruk2022,Dmytruk2021,Dmytruk2024}, elimination of intermediate states~\cite{Arwas2023,Lin2023,Ciuti2021}, and asymptotically decoupling unitary transformation~\cite{Masuki2023,Yuto2022} have been made. 
Exact diagonalization~\cite{Nguyen2024,Macedo2024,Nguyen2023,Gonzalez2023}, density matrix renormalization group~\cite{Baccioni2024,Cordoba2023}, and iterative mean-field approach~\cite{Dmytruk2022,Dmytruk2021} have also been considered to numerically solve the ground state of the full Hamiltonian.

Besides coupling to quantum light, phases of matter can also be altered by disorder. 
The seminal work by P. W. Anderson~\cite{Anderson1958} showed that electronic systems experience a transition from metallic to insulating when disorder is strong enough to localize electrons. In addition to random disorder, localization has also been manifested in other models such as the quasi-periodic Aubry-André model~\cite{Aubry1980}.
In order to undertand the localization-delocalization transition, the Lyapunov and Heurst exponent are studied in one-dimensional systems for diagonal and off-diagonal disorder~\cite{Cheraghchi2005,Gonzalez2019}.
Moreover, disorder can drive the transition from normal insulating phase to topological insulating phase, which is called topological Anderson insulating (TAI) phase~\cite{Li2009}, and the self-consistent Born approximation is used to locate critical disorder strength~\cite{Hsu2020}.
However, little is known about the effect of cavity photons on the TAI phase.

In this paper, we study the longer-range SSH model and the effect of the interplay between cavity photons and disorder.
We introduce the model and the numerical methods in Sec.~\ref{sec:model-hamiltonian}, followed by the energy spectrum of the clean system dressed by photons in Sec.~\ref{sec:photon-band}. 
Sec.~\ref{sec:TAI-cavity} presents the topological phase diagram and localization length of the TAI phase coupled to cavity photons.  
Conclusions are given in Sec.~\ref{sec:conclusion}.

\begin{figure}
\makebox[\textwidth][l]{\includegraphics[width=0.5\textwidth]{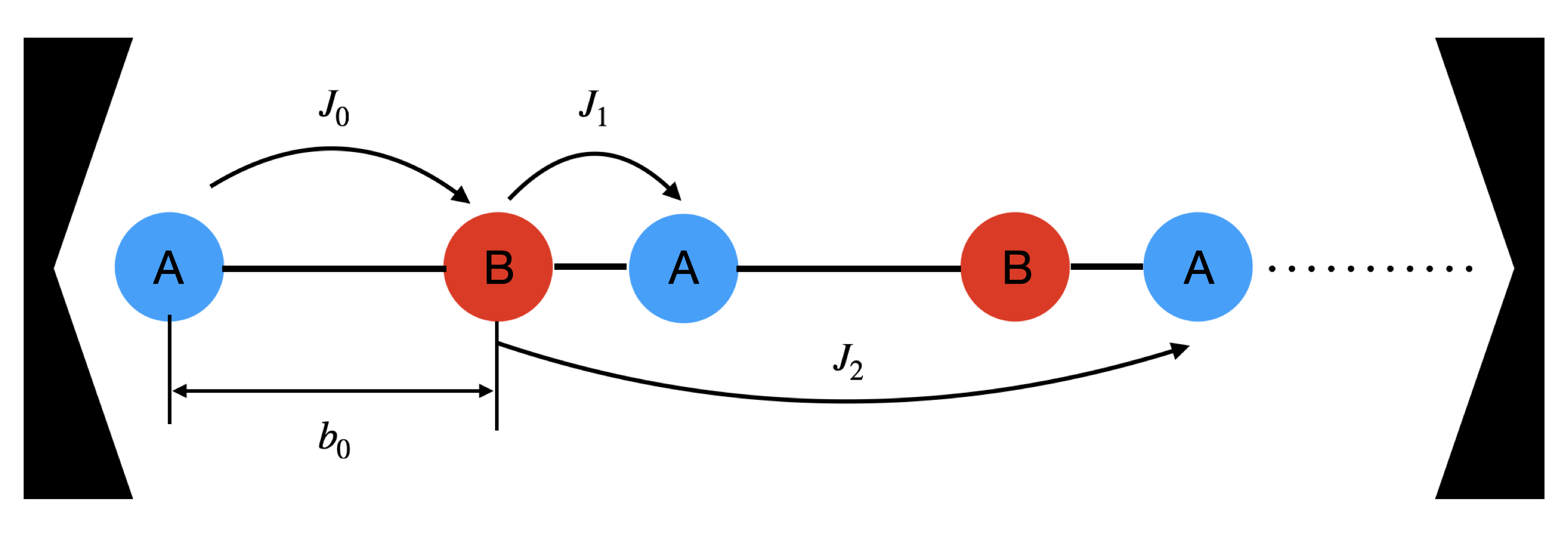}}%
\caption{\label{fig:scheme} Schematic diagram of the system: the longer-range SSH model in a photonic resonator. The blue (red) dots represent $A$ $(B)$ sublattices. The distance between sublattice $A$ and $B$ in the unit cell is denoted by $b_0$. $J_0$, $J_1$ and $J_2$ denote the intracell, intercell and the long-range hopping, respectively. }
\end{figure}

\begin{figure*}
\makebox[\textwidth][c]{\includegraphics[width=\textwidth]{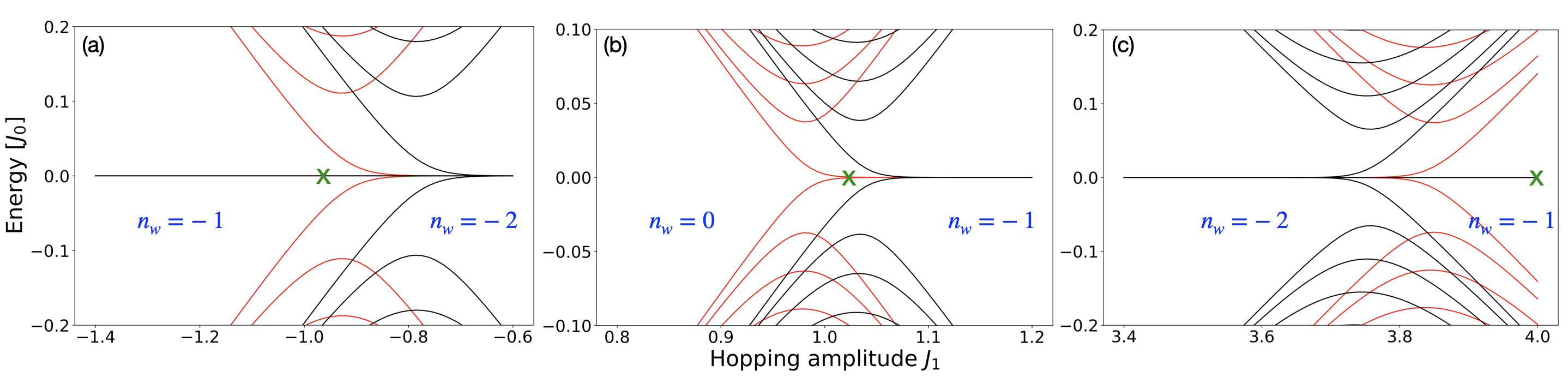}}
\caption{\label{fig:photon-band} Energy spectrum of clean system dressed by photon within regime (\RN{1})-(\RN{3}). (a) $J_2=-2$, (b) $J_2=0$, and (c) $J_2=3$. The $x$-axis is hopping amplitude $J_1$ and $y$-axis is energy. Other parameters are set to $g=6$, $\omega_c=1$, $N_{max}=11$, $L=120$. Intracell distance $b_0$ is set to $0.9$ for red solid line and $0.1$ for black solid line, the green crosses indicate the gap-closing or opening point for $g=0$. The winding numbers are labelled on both sides of the boundary.}
\end{figure*}

\section{\label{sec:model-hamiltonian}Model Hamiltonian}

The bare electronic Hamiltonian is the long-range SSH model that consists of three types of hoppings that preserve chiral symmetry
\begin{eqnarray}\label{eq:ssh}
H_{el}^0&=&J_0\sum_{j=1}^{L}c^{\dagger}_{jA}c_{jB} + J_1\sum_{j=1}^{L-1}c^{\dagger}_{j+1,A}c_{jB} \nonumber\\
&+& J_2\sum_{j=1}^{L-2}c^{\dagger}_{j+2,A}c_{jB} + h.c.,
\end{eqnarray}
where $L$ is the number of unit cell, 
$g$ is the light-matter coupling strength, 
$b_0$ is the sublattice distance, and $J_i$ are hopping amplitudes. The lattice constant is set to $1$. 
The long-range SSH chain is put into a cavity to couple with single mode photons, as shown in Fig.~\ref{fig:scheme}.
Cavity photons couple with electron via Peierls substitution \mbox{\cite{Dmytruk2022,Dmytruk2021}} $J_i\rightarrow J_i e^{i\frac{g}{\sqrt{L}}(a+a^{\dagger})l_i}$
where $l_i$ is the distance between the sites of the corresponding hopping ($l_0=b_0$, $l_1=1-b_0$, $l_2=2-b_0$), and $a$ and $a^{\dagger}$ are the creation and annihilation operator for photons.
The full light-matter Hamiltonian hence reads
\begin{eqnarray}
H=&&J_0\sum_{j=1}^{L}e^{i\frac{g}{\sqrt{L}}b_0(a+a^{\dagger})}c^{\dagger}_{jA}c_{jB} + h.c. \nonumber\\
&&+J_1\sum_{j=1}^{L-1}e^{-i\frac{g}{\sqrt{L}}(1-b_0)(a+a^{\dagger})}c^{\dagger}_{j+1,A}c_{jB} + h.c. \nonumber\\
&&+J_2\sum_{j=1}^{L-2}e^{-i\frac{g}{\sqrt{L}}(2-b_0)(a+a^{\dagger})}c^{\dagger}_{j+2,A}c_{jB} + h.c. \nonumber\\
&&+\omega_c(a^{\dagger}a+\frac{1}{2}),
\end{eqnarray}
with $\omega_c$ being the frequency of cavity photon.
In the numerical calculation, we set $g=6$ and $\omega_c=1$. The choice gives the ratio of light-matter coupling strength and cavity frequency in the same order of magnitude as the experimental setting of the cavity quantum electrodynamics in quantum gas systems~\cite{Cottet2017,Mivehvar2021,Baumann2010}. 
We consider disorder for the intracell hopping, 
\begin{equation}
\label{eqn:HU}
H_U=\sum_{j=1}^{N}U_j e^{i\frac{g}{\sqrt{L}}b_0(a+a^{\dagger})} c^{\dagger}_{jA}c_{jB}+h.c.,
\end{equation}
where $U_j$ are the random disorder distributed uniformly within $\left [ \frac{-U}{2}, \frac{U}{2} \right ]$.

The ground state of the full light-matter Hamiltonian is obtained iteratively under the factorized mean-field ansatz~\cite{Dmytruk2022}, where the full ground state wave function $|\Psi\rangle=|\psi\rangle|\phi\rangle$ is a direct product of the electronic and photonic ground states, denoted by $|\psi\rangle$ and $|\phi\rangle$, respectively. 
First, we solve for the bare electronic ground state of $H_{el}^0$, and use it to obtain the mean-field photonic Hamiltonian $H^{mf}_{ph}=\langle \psi|H|\psi\rangle$. 
Second, we solve for the ground state of $H^{mf}_{ph}$ in the number basis, $|\phi\rangle=\sum_n c_n|n\rangle$, and use it to obtain $H^{mf}_{el}=\langle\phi|H|\phi\rangle$. Third, the ground state of $H^{mf}_{el}$ is solved for and $H^{mf}_{ph}$ is updated.  The second and third step are repeated until convergence of the ground state energy is reached.

\section{\label{sec:photon-band}Energy spectrum in the clean limit}

\begin{figure*}
\makebox[\textwidth][c]{\includegraphics[width=\textwidth]{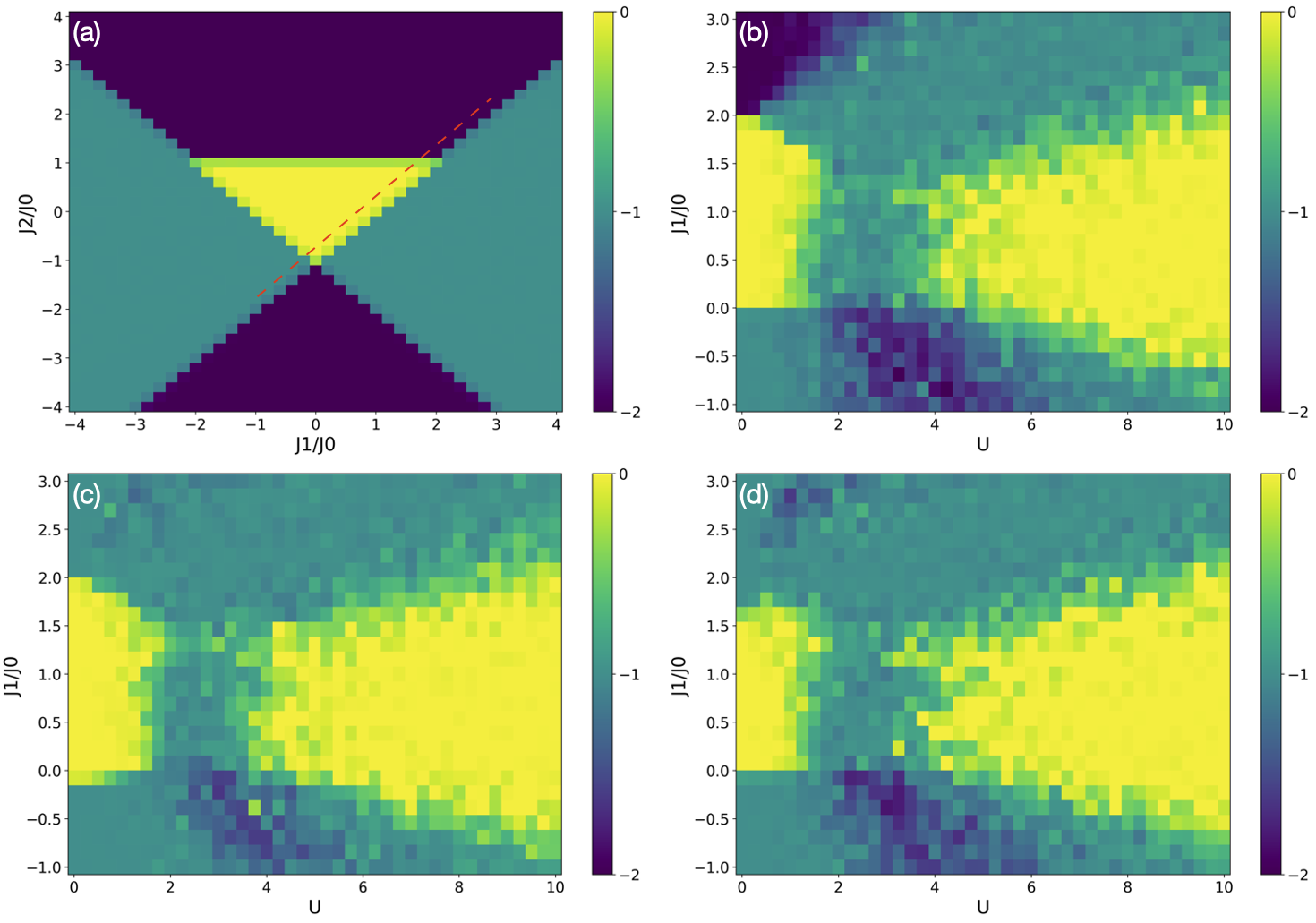}}
\caption{\label{fig:phase} (a) The two-dimensional ($2D$) phase diagram without coupling to photons. Disorder phase diagram along the red dash line in the $2D$ phase diagram for (b) $g=0$ (c) $g=6$, $b_0=0.2$ (d) $g=6$, $b_0=0.8$. In (b-d), the $x$-axis is the disorder strength and the $y$-axis is the hopping amplitude $J_1$ in unit of $J_0$. The number of realization for random disorder is $20$.}
\end{figure*}

For the SSH model, Eq. \ref{eq:ssh}, with open boundary condition, the number of midgap states is twice the winding number, reflecting the system's topological property. Thus, we calculate the energy spectrum to identify the topological phases of the SSH model in a cavity. The results are shown in Fig.~\ref{fig:photon-band}. 
It is found that the effect of the cavity coupling varies with different $J_2$ values. 
When $J_2$ is negative and far from zero, as shown in Fig.~\ref{fig:photon-band}(a), the $n_w=-1$ region is enlarged and the system favors lower winding number, irrespective of $b_0$.
When $J_2$ is close to zero (positive or negative), as shown in Fig.~\ref{fig:photon-band}(b), the $n_w=-1$ region is enlarged for $b_0=0.9$ [red curve], whereas the $n_w=0$ region is enlarged for $b_0=0.1$ [black curve]. Finally, when $J_2$ is positive and far from zero, the $n_w=-1$ region is enlarged for both $b_0=0.9$ and $b_0=0.1$, as shown in Fig.~\ref{fig:photon-band}(c). The system favors lower winding number after coupled to cavity in this regime.
It is worth noting that compared to $b_0=0.1$, $b_0=0.9$ always favor higher winding number, irrespective of $J_2$.
For the sake of convenience in the latter discussion, we label the different $J_2$ regimes represented by Fig.~\ref{fig:photon-band}(a)-~\ref{fig:photon-band}(c) as regime (\RN{1})-(\RN{3}).

The winding number of the bare SSH model is determined by the ratio of the absolute values of the hopping amplitudes. Similarly, 
Fig.~\ref{fig:photon-band} can be explained by the effective hopping amplitudes after the SSH model is dressed by cavity photons. 
Since the mean-field electronic Hamiltonian reads $\langle\phi|H|\phi\rangle$, the effective hopping $\tilde{J}_i$ becomes
\begin{equation}
\tilde{J}_i=J_i\langle\phi|e^{i\frac{g}{\sqrt{L}}l_i(a+a^{\dagger})}|\phi\rangle.
\end{equation}
In the number basis, where $|\phi\rangle=\sum_n c_n|n\rangle$, the numerical results shows a fast decrease in coefficients $c_n$ as $n$ increases.
Thus, the effective hopping is approximated as $\langle 0| e^{i\frac{g}{\sqrt{L}}l_i(a+a^{\dagger})}|0\rangle$. Furthermore, by Taylor expansion of the exponent, the mean-field photonic Hamiltonian and the ground state contain only even powers of the photonic creation and annihilation operators, as a result of the discrete symmetry of the photonic Hamiltonian under $a\rightarrow -a$.  Consequently, only the coefficents $c_n$ with $n\in$ even are nonzero and the effective hopping must be purely  real numbers and 
$\tilde{J}_i$ can be approximated to be $J_i Re\left [ e^{i\mu(l_i)} \right ]$, where the phase angle $\mu(l_i)$ is a real function of $l_i$ that monotonically increases with $l_i$.
Note that the exponent is small for our choice of $g$ and $L$, and the phase angle is expected to be well within 
$\left [ -\pi/2,\pi/2 \right ]$.
In short, cavity photons make hopping more difficult for electrons, and the longer the electron has to jump initially, the harder it is after dressed by photon.

For $J_2\approx0$, in regime(\RN{2}), the topology is decided by $\tilde{J}_1/\tilde{J}_0$ solely. When $b_0>0.5$, the distance $l_1<l_0$ and the phase angle of $\tilde{J}_1$ is smaller than that of $\tilde{J}_0$, resulting in larger $\tilde{J}_1$ which favors $n_w=-1$. 
The opposite is true for $b_0<0.5$.
Away from $J_2=0$ (such as $J_2=3$ or $-2$), the winding number of the system is either $n_w=-1$ or $-2$, depending on $\tilde{J}_2/\tilde{J}_1$. When  $\tilde{J}_2>\tilde{J}_1$, the dressed system has $n_w=-2$. On the other hand, when $\tilde{J}_2<\tilde{J}_1$, the dressed system has $n_w=-1$.

Different from regime (\RN{2}), since $l_2=2-b_0$ and $l_1=1-b_0$, the phase angle of $\tilde{J}_2$ always exceed that of $\tilde{J}_1$, irrespective of the value of $b_0$. 
As a result, the effective long-range hopping $\tilde{J}_2$ is always further reduced compared to $\tilde{J}_1$, and the winding number of the dressed system tends to reduce to $n_w=-1$.
This is manifested by the enlarged $n_w=-1$ region in Fig.~\ref{fig:photon-band}(a) and Fig.~\ref{fig:photon-band}(c).

\section{\label{sec:TAI-cavity}TAI phase in cavity}

In this section, the effect of disorder on the topological phases and the transitions is studied. 
Specifically, we are interested in how cavity photons affect TAI phases. To characterize the topological phases as a function of disorder strength,
 the mean winding number is calculated via the real-space method \cite{Shem2014}
\begin{equation}
w=-\text{Tr}_s \{ Q_{-+}  [ X,Q_{+-}  ]\},
\end{equation}
where $Q_{-+}=S_-QS_+$, $Q_{+-}=S_+QS_-$, $Q=P_+-P_-$, $P_{\pm}$ is projection to positive or negative bands, $S_{\pm}$ is projection to sublattice $A$ or $B$, $X$ is position operator, and $\text{Tr}_s$ is trace over sublattices.

The winding number in the clean limit as a function of the hopping strengths is shown in Fig.~\ref{fig:phase}(a).
To study TAI phase, we chose the hopping amplitudes along a path that is parallel to the bare electronic phase boundary with a small offset [the red dashed line in Fig.~\ref{fig:phase}(a)]. 
Fig.~\ref{fig:phase}(b) examines the disorder induced phase transitions without cavity photons.
Figs.~\ref{fig:phase} (c) and (d) include the light-matter coupling for different sublattice distance $b_0=0.2$ and $b_0=0.8$, respectively.
Previous studies on the disorder effects in the SSH or bipartite chain~\cite{Gonzalez2019,Hsu2020,Inui1992} have shown that the midgap states remain stable until disorder strength is comparable to the size of bandgap.
Thus, we use the band gap to estimate the critical disorder strength at the disorder-induced phase transitions. 
The modulation of the band gap is controlled by the effective intercell hopping $\tilde{J}_i$  and the dressed intracell hopping $\tilde{J}_0$. In addition, we note that the hopping disorder is dressed by the same amount as the corresponding $J_0$, as suggested by Eq.~(\ref{eqn:HU}).

\begin{figure}
\makebox[\textwidth][l]{\includegraphics[width=0.5\textwidth]{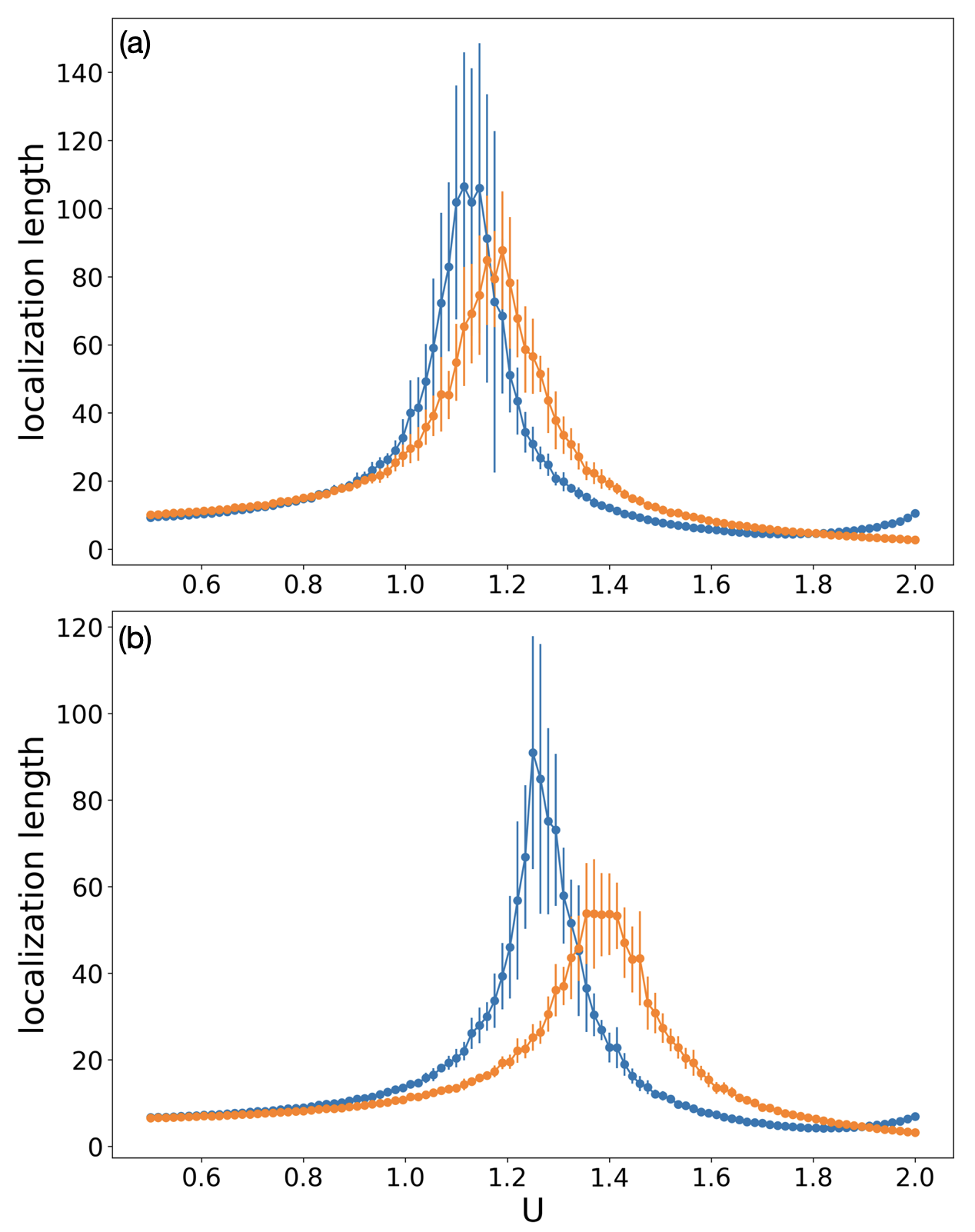}}
\caption{\label{fig:ll} Localization length along $J_1/J_0=0.1$ (blue) and $J_1/J_0=-0.1$ (orange) for (a) $g=0$, the same parameters as in Fig.~\ref{fig:phase}(b), and (b) $g=6, b_0=0.8$, the same parameters as in Fig.~\ref{fig:phase}(d). The peaks indicate phase transition point while the other regions denote localization. The solid points are averages over $20$ disordered realizations.}
\end{figure}

For regime (\RN{1}), the critical disorder strength is increased for the region near $J_1=-1$ in Fig.~\ref{fig:phase}(c-d), compared to Fig.~\ref{fig:phase}(b). 
This can be understood as the enlarged bandgap after coupling to cavity photons in the clean limit, as shown in Fig.~\ref{fig:photon-band}(a).
Within regime (\RN{1}) however, the shift in critical disorder strength varies for different $J_1$.  
Specifically, a comparison between Fig.~\ref{fig:phase}(b) and Fig.~\ref{fig:phase}(d) at  $J_1=\pm0.1$ shows that the transition point is almost unchanged for $J_1=0.1$ while the critical disorder strength is increased for $J_1=-0.1$.
In Fig.~\ref{fig:phase}(b), there are two different phases ($n_w=-1$ and $n_w=-2$) along the fixed disorder strength at $U=2$ across $J_1=0$, while in Fig.~\ref{fig:phase}(d) the system stays in the same phase ($n_w=-1$) across $J_1=0$

For regime (\RN{2}), the photon dressed bandgap is increased for $b_0<0.5$ and reduced for $b_0>0.5$ [see Fig.~\ref{fig:photon-band}(b)].
Since the effective disorder strength is reduced by cavity photons, the resulting critical disorder strength is determined by a competition between the reduced bandgap and the reduced hopping disorder. 
A close inspection between Fig.~\ref{fig:phase}(b) and Fig.~\ref{fig:phase}(d) around $1<J_1<1.5$ shows the decrease of critical disorder strength, which implies that the decrease in bandgap has a larger impact than the reduced hopping disorder. 

For regime (\RN{3}), the bandgap is enlarged for both $b_0$ [see Fig.~\ref{fig:photon-band}(c)].
Moreover, the winding number between the $g=0$ and $g\neq0$ systems is different in the clean limit.
Consequently, the absence of the $n_w=-2$ region is observed on the top left corner of Fig.~\ref{fig:phase}(c) and Fig.~\ref{fig:phase}(d).
In this regime, the bandgap is enlarged more significantly for $b_0<0.5$, making the TAI transition from $n_w=-1$ to $n_w=-2$ nearly absent in Fig.~\ref{fig:phase}(c).

To ensure localization in trivial and topological insulating phases, and to better locate the phase transition boundary, the localization length is calculated by the iterative Green’s function method \cite{MacKinnon1983}
\begin{equation}
\frac{2}{\lambda}=-\lim_{L \to \infty}\frac{1}{L}\text{Tr ln}|G_{1L}|^2,
\end{equation}
where $G_{1L}$ is the Green’s function connecting the first and last slice of the system. The renormalized $\tilde{J}_i$ were calculated with a finite chain size ($L=120$) and applied to the slice Hamiltonian to compute $\lambda$ with $L=2\cdot10^4$.
The localization lengths for $g=0$ and $g=6$ with $b_0=0.8$ are shown in ~\ref{fig:ll} (a) and (b), respectively. The blue (orange) curve in the figure stands for the localization lengths for $J_1=0.1, \, (-0.1)$. The strong peaks identify topological phase transitions. 
It is shown in Fig.~\ref{fig:ll}(a) that the two peaks are close for the bare SSH model. 
After adding cavity photons, the blue peak remains mostly unchanged, while the orange peak is shifted to the right in Fig.~\ref{fig:ll}(b). The shift of the peaks in localization length for $J_1=-0.1$ supports the observation in Fig. \ref{fig:phase} (d) that the critical disorder strength increases when the system is coupled with cavity photons.

\begin{figure}
\makebox[\textwidth][l]{\includegraphics[width=0.5\textwidth]{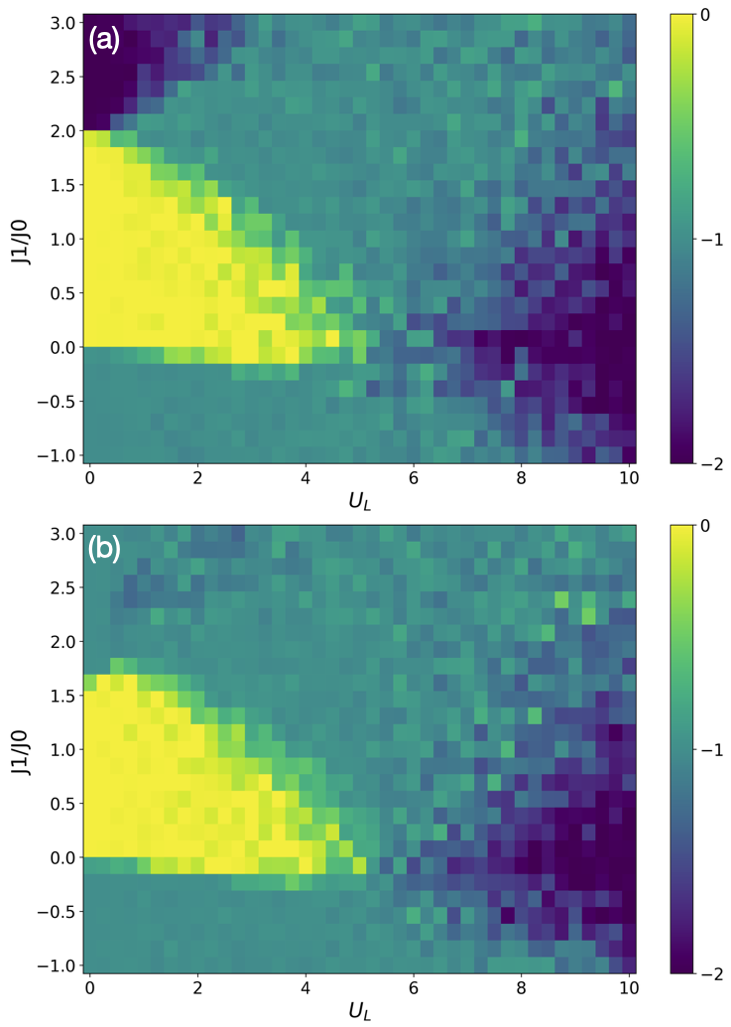}}
\caption{\label{fig:ul} TAI phase for longer-range disorder $U_L$ drawn along $J_2=J_1-0.94$ for (a)$g=0$ (b)$g=6$, $b_0=0.8$. The number of disorder realizations is $20$.}
\end{figure}

\section{\label{sec:conclusion}Conclusion}
In summary, we found that cavity photon can modify hopping amplitudes, resulting in the change of phase transition boundaries. 
Depending on the value of $J_2$, coupling to photon can either affect the topology differently for $b_0<0.5$ and $b_0>0.5$ or merely lower the winding number of the system. 
When disorder is introduced, it could drive transition to TAI phase where the critical disorder depends on the size of the photon-dressed bandgap and the reduced disorder strength. 
Since the effect of cavity photons varies for different $b_0$ and $J_2$, the critical disorder strength can hence be modulated.
Localization length is calculated to ensure localization away from phase boundaries and to better locate phase transitions.

\begin{acknowledgments}
Y.-C. Shaw, H. C. Hsu and J. S. You are supported by the National Science and Technology Council (Grant No. 110-2112-M-003-008-MY3, 113-2628-M-004-001-MY3). 
The authors acknowledge Academia Sinica Grid Computing Centre, Grant No. AS-CFII-112-103, for the computation resources.
H. C. Hsu and J. S. You are also grateful for support from National Center for Theoretical Sciences in Taiwan.
Y.-C. Shaw thanks Tian-Shu Guo for helpful discussions.
\end{acknowledgments}

\appendix

\section{TAI phase for longer-range disorder}
To consider the effect of longer-range disorder, Eq.~\ref{eqn:HU} should be modified to 
\begin{equation}
\label{eqn:HUL}
H_{U_L}=\sum_{j=1}^{N}{U_L}_j e^{i\frac{g}{\sqrt{L}}(2-b_0)(a+a^{\dagger})} c^{\dagger}_{j+2,A}c_{jB}+h.c.,
\end{equation}
where ${U_L}_j$ is the random disorder on the longer-range hopping distributed uniformly within $\left [ \frac{-U_L}{2}, \frac{U_L}{2} \right ]$.
The results are presented in Fig.~\ref{fig:ul}, where except from region near $U_L= 0$, the effect of cavity photons is not obvious.
Both Fig.~\ref{fig:ul}(a) and Fig.~\ref{fig:ul}(b) show similar critical disorder strength at transitions from $n_w=0$ to $n_w=-1$ and $n_w=-1$ to $n_w=-2$.


\end{document}